# Reduced-Graphene-Oxide with Traces of Iridium or Gold as Active Support for Pt Catalyst at Low Loading during Oxygen Electroreduction

P. J. Kulesza[a], B. Dembinska[a], S. Zoladek[a], I. A. Rutkowska[a], K. Miecznikowski[a], E. Negro[b], V. Di Noto[b]

[a]Department of Chemistry, University of Warsaw, Pasteura 1, PL-02-093 Warsaw, Poland
[b]Department of Industrial Engineering, Università degli Studi di Padova in Department of Chemical Sciences, Via Marzolo 1, 35131 Padova (PD) Italy

Chemically-reduced graphene-oxide-supported gold or iridium nanoparticles are considered here as active carriers for dispersed platinum with an ultimate goal of producing improved catalysts for electroreduction of oxygen in acid medium. Comparison is made to the analogous systems not utilizing reduced graphene oxide. High electrocatalytic activity of platinum (loading up to 30 µg cm$^{-2}$) dispersed over the reduced-graphene oxide-supported Au (up to 30 µg cm$^{-2}$) or Ir (up to 1.5 µg cm$^{-2}$) nanoparticles toward reduction of oxygen has been demonstrated using cyclic and rotating ring-disk electrode (RRDE) voltammetric experiments. Among important issues are possible activating interactions between gold and the support, as well as presence of structural defects existing on poorly organized graphitic structure of reduced graphene oxide. The RRDE data are consistent with decreased formation of hydrogen peroxide.

## Introduction

There has been growing interest in the field of oxygen electroreduction, particularly with respect to potential applications in the science and technology of low-temperature fuel cells (1-11). Obviously, many efforts have been made to develop suitable alternative electrocatalysts efficient enough to replace electrocatalysts based on scarce strategic elements such as platinum-group metals (5,6,8,9,12,13). Despite intensive research in the area, there are still a number of fundamental problems to be resolved, and the practical oxygen reduction catalysts still utilize systems based on platinum.

The $O_2$-reduction electrocatalysts are typically nanocomposite materials utilizing metal nanoparticles bearing the active sites dispersed on suitable supports. While exhibiting long-term stability, a useful support should facilitate dispersion, provide easy access of reactants, and assure good electrical contact with active sites. In spite of limitations related to the durability, carbon nanoparticles of approximately 20-50 nm diameters (e.g. Vulcan XC-72R) are commonly utilized as supporting materials. Because of the high specific surface area and excellent thermal, mechanical and electrical properties, graphene and graphene-based materials (14-16) have recently been considered as supports for catalysts (17-21). Under such conditions, the parasite effects related to agglomeration and thus degradation of catalytic nanoparticles are likely to be largely prevented.

In the present work, we consider the chemically-reduced-graphene-oxide-supported gold or iridium nanostructures as carriers for dispersed Pt nanoparticles as catalytic systems for the electroreduction of oxygen in acid medium (0.5 mol dm$^{-3}$ $H_2SO_4$). Among important issues is the ability of the proposed carriers to act as the systems effectively inducing decomposition of the hydrogen peroxide undesirable intermediate (5,22). The latter problem is expected to become an issue when the catalytic platinum would be utilized at low loadings. Here, we propose to decorate the graphene based carriers with gold nanoparticles (loading, 30 µg cm$^{-2}$) or with traces amounts of iridium (loading, 1-2 µg cm$^{-2}$). The usefulness of Au nanostructures during the reduction of oxygen has been recently demonstrated (23). Here application of inorganic Keggin-type heteropolymolybdates ($PMo_{12}O_{40}^{3-}$) as capping ligands (capable of chemisorbing on both gold and carbon substrates (24-31)) facilitates deposition, nucleation, stabilization and thus controlled growth of gold nanoparticles on surfaces of both Vulcan and graphene nanostructures. Furthermore, we have utilized the so-called reduced graphene oxide which, contrary to conventional graphene, still contains oxygen functional groups regardless of subjecting it to the chemical reduction step (14-16). By analogy to graphene oxide, the existence of oxygen groups in the plane of carbon atoms of reduced graphene oxide not only tends to increase the interlayer distance but also makes the layers somewhat hydrophilic. Furthermore, during fabrication of the catalytic systems (in acid medium), the adsorbed polymolybdates (9,25) are likely to bind gold nanoparticles via the oxygen or hydroxyl groups on graphene and Vulcan surfaces. Finally, we explore here the reduced graphene oxide based carriers decorated with catalytic iridium. It is noteworthy that iridium, even at trace levels, has been found to exhibit high reactivity toward the reductive decomposition of hydrogen peroxide (32). As a rule, the electrocatalytic diagnostic experiments described herein involve comparative measurements utilizing commonly-used Vulcan (carbon) supports as carriers for Pt nanoparticles deposited at the same loadings (typically 15-30 µg cm$^{-2}$) as in the case of hybrid systems with the reduced graphene oxide. It is apparent from the diagnostic cyclic voltammetric and rotating ring-disk measurements that the systems utilizing the reduced-graphene-oxide-supports decorated with Au-nanoparticles or traces of iridium could act as active matrices for Pt catalysts thus forming the potent $O_2$-reduction electrocatalytic systems. In particular, the proposed systems have exhibited higher electrocatalytic currents and produced lower amounts of the undesirable hydrogen peroxide intermediate during oxygen reduction. The enhancement effect is particularly found in the high potential range (0.8-1.0 V vs. RHE). On the whole, the combined effect of the high surface area and electrical conductivity of reduced graphene oxide should also contribute to the overall enhancement effect.

### Experimental

All chemicals were analytical grade materials and were used as received. Solutions were prepared from the distilled and subsequently deionized water. They were deoxygenated by bubbling with ultrahigh purified nitrogen. Experiments were carried out at room temperature (22±2 °C).

The 5% Nafion-1100 solution was purchased from Aldrich. Platinum black nanoparticles were obtained from Alfa Aesar. Sulfuric acid was from POCH (Poland). Graphene oxide sheets of 300-700 nm sizes (thickness, 1.1±0.2 nm) were from Megantech. Reduced graphene oxide (rGO) was obtained using sodium borohydride as reducing agent at 80°C according to the procedure described earlier (33).

The syntheses of phosphomolybdate-modified gold nanoparticles supported onto Vulcan XC72R carbon and reduced graphene oxide matrices were performed in the analogous manner as described earlier (26-30) but in the presence of an appropriate carbon support. The stoichiometric volume of freshly prepared aqueous sodium tetrahydroborate ($NaBH_4$) was added to the phosphomolybdate-functionalized carbon supports in order to transform the oxidized $H_3PMo_{12}O_{40}$ adsorbates into partially reduced $H_3[H_4P(Mo^V)_4(Mo^{VI})_8O_{40}]$ heteropolyblue forms. To obtain a gold loading on the level of 30 wt% of Au on the appropriate heteropolyblue-modifed carbon, an equivalent volume of the aqueous 7.5 mmol dm$^{-3}$ chlorauric acid ($HAuCl_4$) solution was added to the respective suspension. As a rule, appropriate amounts of the resulting catalytic inks were dropped onto surfaces of glassy carbon electrodes to obtain loadings of gold nanoparticles equal to 30 μg cm$^{-2}$.

Electrode layers were deposited on glassy carbon disk electrodes by introducing (by dropping) appropriate volumes of inks containing catalytic nanoparticles and using 2–propanol and Nafion® (20% by weight) as solvent and binder, respectively.

Carbon Nanotube (CNT) supported Pt (20%w) and $SiO_2$-doted reduced-graphene-oxide (rGO-$SiO_2$) supported iridium Ir (2%w) were obtained through thermal reduction (under reflux, Ar atmosphere) of $H_2PtCl_6$ and $IrCl_3$, respectively, in the presence of an appropriate support. Before electrochemical examination, the hybrid system was produced by subjecting Pt20%-CNTs to grinding (1:1, by weight) with Ir2%-rGO-$SiO_2$. Catalytic inks were prepared with 2-propanol and Nafion® (10%w).

All electrochemical measurements were performed using CH Instruments (Austin, TX, USA) 760D workstations in three electrode configuration. The reference electrode was the $K_2SO_4$-saturated $Hg_2SO_4$ electrode, and a carbon rod was used as a counter electrode. As a rule the potentials reported here were recalculated and expressed vs. Reversible Hydrogen Electrode (RHE). Glassy carbon disk (geometric area, 0.071 cm$^2$) working electrodes were utilized as substrates. The rotating ring disk electrode (RRDE) working assembly was from Pine Instruments; it included a glassy carbon (GC) disk and a Pt ring. The radius of the GC disk electrode was 2.5 mm; and the inner and outer radii of the ring electrode were 3.25 and 3.75 mm, respectively.

Morphology of samples was assessed using Libra Transmission Electron Microscopy120 EFTEM (Carl Zeiss) operating at 120 kV. The Raman spectra were collected with a confocal Raman Microscope (model DRX, Thermo Scientific) and using an excitation laser with a wavelength of 532 nm.

**Results and Discussion**

Physicochemical Identity of Graphene Nanostructures

Graphene oxide, GO, and partially reduced graphene oxide, rGO contain various carbon–oxygen groups (hydroxyl, epoxy, carbonyl, carboxyl), in addition to the large population of water molecules still remaining in the reduced samples. Independent elemental analysis based on the C 1s and O 1s XPS spectra from XPS measurements showed that the oxygen content in rGO was in the range from 8.6 to 12.1 at%; the C-to-O ratio was on the level of 7.1–10.3. When compared to the analogous parameters of the commercially available GO, the oxygen content and the C-to-O ratio values were more than three times lower and more than three times higher, respectively. Furthermore, the Raman spectra show two large peaks in the range of 1300-1600 cm$^{-1}$: one peak near 1350 cm$^{-1}$, which stands for the D band originating from the amorphous structures of carbon, and the second one close 1580 cm$^{-1}$, which is correlated with the G band, reflects the graphitic structures of carbon. Simple comparison of intensities of G and D bands in rGO, relative to the analogous bands in GO, are lower and higher, respectively. This result implies presence of interfacial defects as well as the lower degree of organization of the graphitic structure of rGO relative to GO.

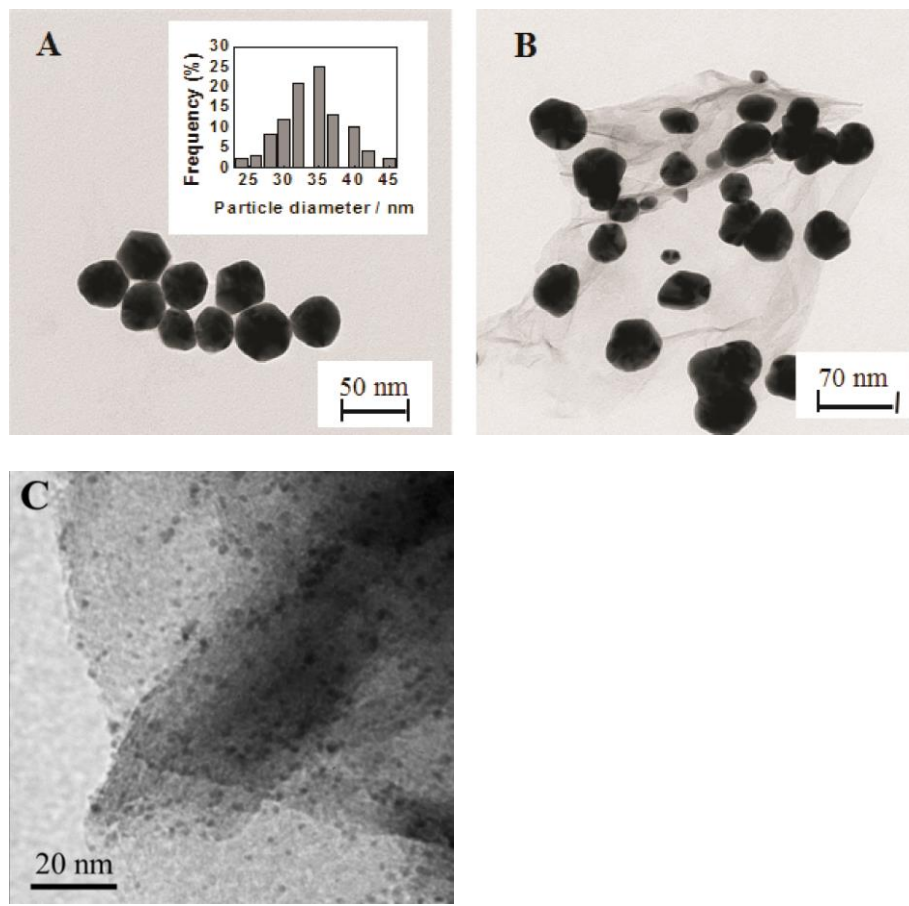

Figure 1. Transmission electron microscopic images of (A) unsupported gold nanoparticles, (B) reduced-graphene-oxide (rGO) supported Au nanoparticles, and (C) rGO-SiO$_2$-supported iridium nanoparticles.

The phosphomolybdate (H$_3$PMo$_{12}$O$_{40}$) modified Au nanoparticles (supported and unsupported) were characterized using Transmission Electron Microscopy (TEM). It is apparent from Figure 1 that, while unsupported gold particles (Figure 1A) have diameters that are fairly uniform in the range between *ca.* 30 and 40 nm diameters, the rGO-supported particles (although slightly larger and less uniform) have comparable sizes typically ranging from 30 to 50 nm (Figure 1B). On mechanistic grounds, gold nucleation may occur at the rGO "defect" sites, including surface polar groups and polyoxometallate adsorbates. It is reasonable to expect that the partially reduced (heteropolyblue) PMo$_{12}$O$_{40}^{3-}$ sites induce generation of somewhat larger gold nanoparticles.

Figure 1C illustrates TEM of iridium nanoparticles generated onto rGO-SiO$_2$ support. Among important issues is the low size (less than 2 nm) and the intended very low loading (<2 µg cm$^{-2}$).

Reduction of O$_2$ at Pt Nanoparticles Deposited onto rGO-supported Au

The rGO-supported Au nanoparticles are obviously less active than conventional Vulcan-supported Pt during electroreduction of oxygen under conditions of the RRDE voltammetric diagnostic experiments at the comparable loadings (30 µg cm$^{-2}$). In the present work, we also disperse (onto rGO-supported Au) bare Pt nanoparticles (sizes 7-8 nm). But comparison to the commercially available Vulcan-supported Pt is not straightforward here because such carbon-supported Pt nanoparticles have sizes on the level of a few (3-4) nm whereas our Au nanoparticles are much larger (40-50 nm diameters). As mentioned above, we have dispersed the commercially available unsupported Pt nanoparticles (sizes 7-8 nm) at the loadings of 30 μg cm$^{-2}$ over two different supports or catalytic systems considered here: (a) simple (bare) gold nanoparticles, and (b) reduced graphene oxide (rGO) supported gold nanoparticles. It is clear from the RRDE experiments (Figure 2A) that the disk currents have occurred to be somewhat higher during the reduction of oxygen at Pt nanoparticles deposited onto the rGO-supported-gold nanostructures relative the performance of Pt deposited onto bare gold nanoaprticles. In the case of Pt nanoparticles deposited onto the rGO-supported-gold, the ring currents (Figure 2B) have been also lower what is consistent with the less pronounced formation of hydrogen peroxide. Figure 2C illustrates the percent amount of $H_2O_2$ (%$H_2O_2$) formed during reduction of oxygen under the conditions of RRDE voltammetric experiments of Figure 2A. The actual calculations have been done using the equation given below:

$$\%_{H2O2} = 200 * I_{ring}/N / (I_{disk} + I_{ring}/N) \qquad [1]$$

where $I_{ring}$ and $I_{disk}$ are the ring and disk currents, respectively, and N is the collection efficiency (equal to 0.39). The results clearly show that the production of $H_2O_2$ is the lowest for system utilizing gold nanoparticles supported onto chemically-reduced graphene-oxide.

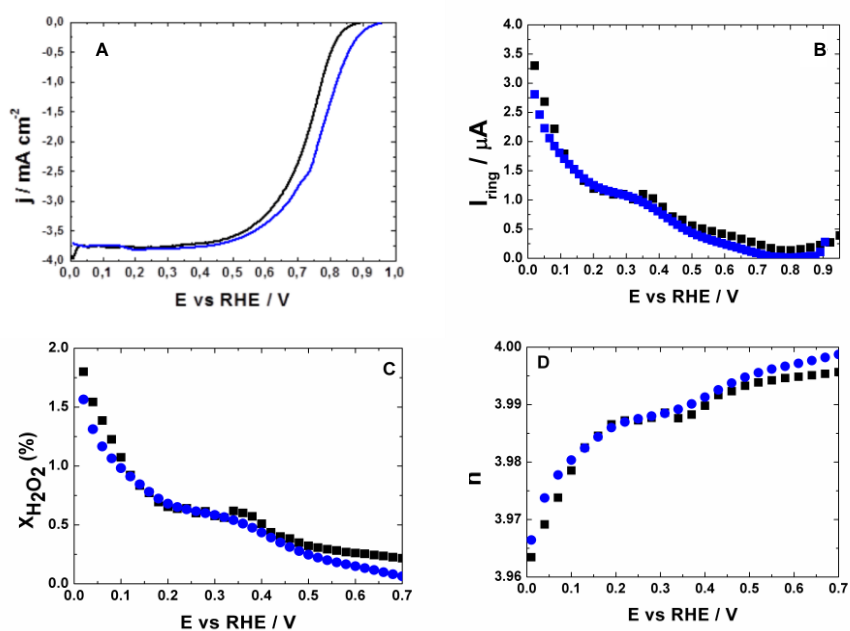

Figure 2. (A) Normalized (background subtracted) rotating ring-disk voltammograms for oxygen reduction at Pt nanoparticles dispersed over the unsupported gold nanostructures (black line), and the rGO-supported gold nanostructures (blue line). (B) Ring currents are recorded upon application of 1.2 V. (C) Percent fraction of hydrogen peroxide produced during electroreduction of oxygen (and detected at ring). (D) Number of transferred electrons (n) per $O_2$ molecule. Electrolyte: $O_2$-saturated 0.5 mol dm$^{-3}$ $H_2SO_4$. Scan rate: 10 mV s$^{-1}$. Rotation rate: 1600 rpm.

The overall number of electrons exchanged per $O_2$ molecule (n) was calculated as a function of the potential using the RRDE voltammetric data of Figure 2A and B and using the equation mentioned below:

$$n = 4 * I_{disk} / (I_{disk} + I_{ring}/N) \qquad [2]$$

The corresponding number of transferred electrons (n) per oxygen molecule (Figure 2D) involved in the oxygen reduction was obviously higher in a case of the system utilizing Au nanoparticles supported onto rGO.

Reduction of $O_2$ at Hybrid Catalyst of Pt20%-CNTs Admixed with Ir2%-rGO-SiO$_2$

Figure 3 illustrates representative (A) disk (voltammetric) and simultaneous (B) ring (upon application of 1.2 V) steady-state currents recorded during the reduction of oxygen (in the $O_2$-saturated 0.5 mol dm$^{-3}$ H$_2$SO$_4$ at 1600 rpm rotation rate and 10 mV s$^{-1}$ scan rate) using the hybrid catalyst composed of Pt20%-CNTs admixed (at 1:1 ratio) with Ir2%-rGO-SiO$_2$ (red line) and the analogous Ir-free system (black line). It is noteworthy that loadings of Pt and Ir are on the levels 15 and 1.5 µg cm$^{-2}$, respectively. Under hydrodynamic voltammetric conditions of Figure 3, while the disk current densities are roughly comparable for all Pt-containing systems (Figure 3A), the different ring currents (Figure 3B) have been produced clearly implying formation of lower amounts of the undesirable H$_2$O$_2$ intermediate in a case of the system containing traces of iridium. Furthermore, when the percent values for the hydrogen peroxide intermediate formation are compared (Figure 3C), it becomes apparent that they are particularly low (below 1%) in the presence of the traces of Ir (rGO-SiO$_2$-supported) at positive potentials (0.6-0.9 V vs. RHE). Finally, the electroreduction of oxygen proceeds now at more positive potentials (Figure 3A) in spite of the low Pt-loading and the ultra-low addition of Ir. The above observations are of potential importance to the development of catalytic systems for low temperature fuel cells.

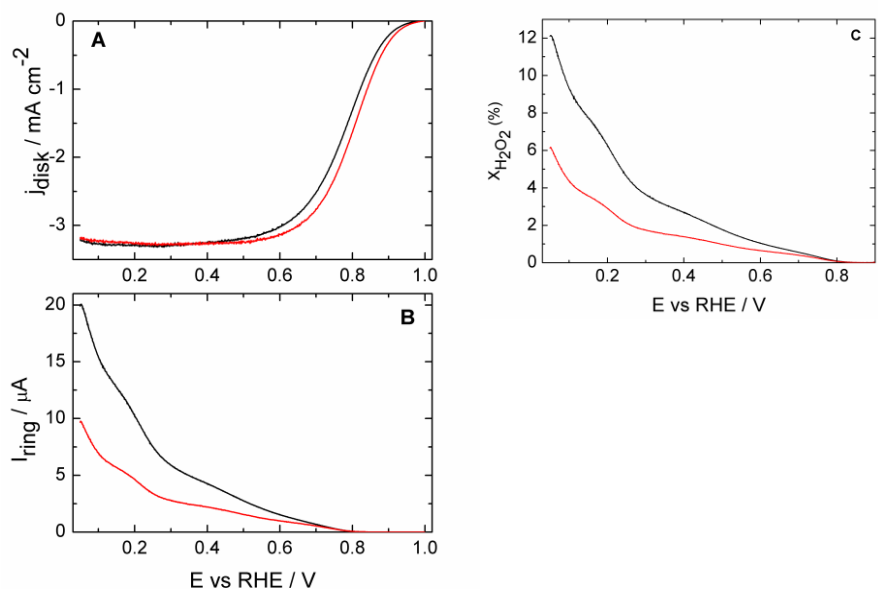

Figure 3. (A) Rotating ring disk voltammograms for the reduction of oxygen at CNT-supported Pt admixed with SiO$_2$-doted-rGO-supported Ir (red line) and the analogous Ir-free system (black line). (B) H$_2$O$_2$ oxidation currents recorded at the Pt ring upon application of 1.2 V. (C) Percent of formation of H$_2$O$_2$. Other conditions as for Figure 2.

## Conclusions

This study clearly demonstrates that the chemically-reduced graphene-oxide, while decorated with gold or iridium nanostructures, acts as a robust and activating support for dispersed Pt nanoparticles during electrocatalytic reduction of oxygen in acid medium (0.5 mol dm$^{-3}$ $H_2SO_4$). For the same loading of catalytic gold nanoparticles (30 µg cm$^{-2}$), application of the reduced graphene oxide support results in formation of lower amounts of the undesirable $H_2O_2$ intermediate. Moreover the onset potential for the oxygen reduction has been the most positive (0.9 V) in a case of the system utilizing reduced graphene oxide. Synergistic effects and activating interactions between catalytic metal nanoparticles and nanostructured graphene supports cannot be excluded here with respect to lowering the dissociation activation energy for molecular $O_2$ through accelerating the charge transfer from metal in the presence of graphene and by reducing stability of the $H_2O_2$ intermediate species.

We have also demonstrated here that co-existence of the carbon-nanotube-supported Pt nanoparticles and the reduced graphene oxide supported iridium nanostructures at low loadings (15 and 1.5 µg cm$^{-2}$, respectively) yields highly active electrocatalytic system for the electroreduction of oxygen in acid medium. The enhancement effect coming from the addition of traces of iridium (supported onto the silica doped reduced graphene oxide) may originate from the high ability of Ir to induce decomposition of the undesirable hydrogen peroxide intermediate. The presence of carbon nanotubes may improve charge distribution at the electrocatalytic interface. Further research is needed to elucidate possible specific interactions.

Our preliminary results with platinum dispersed over the reduced-graphene-oxide-supported gold or iridium imply that such a hybrid catalyst (once optimized with respect to minimizing of particle sizes and loadings of Pt) could be of interest in the fuel cell science and technology.

## Acknowledgements

This work was supported by the European Commission through the Graphene Flagship – Core 1 Project (GA-696656). The Polish side appreciates support from National Science Center (Poland) under Maestro Project 2012/04/A/ST4/00287.